\documentclass{webofc}
\usepackage[varg]{txfonts}   
\usepackage{slashed} 
\begin{document}
\title{QCD strings and the U(1) problem}
%
%

\author{\firstname{CHI} \lastname{XIONG}\inst{1,2}\fnsep\thanks{\email{xiongchi@ntu.edu.sg}} 
}

\institute{Institute of Advanced Studies,  Nanyang Technological University, Singapore
\and
           School of Physical and Mathematical Sciences,  Nanyang Technological University, Singapore
          }

\abstract{%
We promote the usual QCD $\theta$-parameter to a field and interpret it as the phase of the quark condensate, which becomes nontrivial when topological defects, vortices in our formulation, are induced in the quark condensate by the QCD strings (chromoelectric flux tubes). The QCD topological term emerges naturally as a derivative coupling between the Chern-Simons current and a supercurrent in the quark condensate. This new formulation can address the $U_A(1)$ problem and leads to the chiral magnetic effects. It is possible that in this formulation the strong CP problem can be avoided without the axion particle.
}
\maketitle
%
\section{Introduction}
\label{intro}

The $U_A(1)$ problem, i.e. the absence of a ninth light pseudoscalar meson or the reason why the $\eta'$ meson is much heavier than expected, was considered a solved problem long time ago. Firstly Kogut and Susskind noted the necessary ingredient for the solution should be a pole at $q^2 =0$ in the matrix elements of some gauge-invariant operators \cite{KS74}; 't Hooft then noticed that instantons can lead to an explicit solution as a result of the axial anomaly \cite{tHooft76}, and  Crewther studied the problem using anomalous Ward identities \cite{Crewther}; Witten explained this problem from the large $N_c$ point of  view \cite{Witten}and more concretely, Veneziano introduced a ghost state and showed the possibility to calculate the mass of the $\eta'$ meson \cite{Veneziano}.  Nevertheless since all these solutions involve a topological charge term (e.g. the usual $\theta$-term) in quantum chromodynamics (QCD), which cause another problem of the CP violation in the strong interaction (the strong CP problem), it seems to be still interesting to consider other scenarios to address the $U_A(1)$ problem again.

In this presentation we review a different approach to study this problem based on a QCD string (flux-tube) model. The new ingredients of this approach is that during the course of solving the $U_A(1)$ problem we try to keep the confining feature of the QCD vacuum and try to avoid the strong CP problem as well.
The QCD string is constructed based on a decomposition of the gluon field  (by Cho \cite{Cho}, Duan and Ge \cite{DuanGe}, and Faddeev and Niemi \cite{Faddeev}. See Ref. \cite{Kondo14} for a review). We also introduce an order parameter for the quark condensate which is treated a complex scalar field. An effective Lagrangian is given to describe the string (flux tube) configuration consisting of the abelianized gluon potential,  the order parameter for the quark condensate, the quarks and meson channels (scalar and pseudoscalar). Our focus is the QCD topological charge term and we demonstrate that, assuming quantum vorticity in the quark condensate,  a topological term can naturally emerge as an effective action in the bulk space due to the anomaly-inflow mechanism discovered by Callan and Harvey \cite{Callan-Harvey}. Remarkably, the role of the usual QCD $\theta$-parameter is played by the phase of the quark condensate and the $U_A(1)$ problem can be solved in a similar way. We use the large-$N_c$ method and the 't Hooft effective vertex to describe briefly how it can be done. Other applications in topics such as chiral magnetic effect and the possibility of avoiding the strong CP problem are discussed as well.

\section{Quark condensate as order parameter}
\label{sec-1}

The first new ingredient of our formulation is a complex order parameter for the quark condensates.  
We will work within the SU(2) QCD and two flavors ($N_f = 2$) for simplicity and introduce a complex scalar  
\begin{equation} 
\phi (x) \equiv \textrm{Re}\phi (x) + i \, \textrm{Im}\phi (x) \equiv f (x) \,e^{i \alpha (x)} 
\end{equation}
where the functions $f (x)$ and $\alpha (x)$ are the magnitude and phase of $\phi$, respectively. They are related to the quark condensate as
\begin{equation}
\left\langle \bar{\psi}^i_R \psi^j_L \right\rangle = - \phi (x) \, \delta^{ij} = - f (x) \,e^{i \alpha (x)} \, \delta^{ij}
\end{equation} 
Note that here we do NOT make the assumption that the QCD vacuum has a definite parity, hence in general $\phi$ is complex, similar to the cases of Bose-Einstein condensation, superfluidity and superconductivity where the order parameters are complex functions (e.g. the gap function in low-temperature superconductivity theories). It is easy to see that $\phi$ does transform as a complex field under the $U_A(1)$ symmetry. We will show later the importance of introducing a non-vanishing phase field $\alpha = \alpha(x)$. For simplicity we also take one value for different quark condensates while in general we should treat the $\phi$ field as a matrix. 

\section{QCD string/flux tube via an abelian decomposition of gluon potential}
\label{sec-2}

QCD strings (chromoelectric flux tubes) can be generated by quark pair creation, e.g. in the process of hadron production in $e^{+}e^{-}$ annihilation \cite{Casher}. Given enough energy, a quark pair $q\bar{q}$ is created at some point and then quark and anti-quark move away in opposite direction at a speed $v$ close to the speed of light. The chromoelectric flux tube is then formed when the distance between the quark pair becomes $\sim 1 \, \textrm{GeV}^{-1}$.
The formation of quark-gluon plasma may also involve color flux tubes. During the ultra-relativistic collision of heavy nuclei, net color charges develop and consequently color flux tubes connect these color charges. Quark-antiquark pairs are produced from the color fields via the Schwinger mechanism. In our formulation these QCD strings bring inhomogeneity to the quark condensate and induce quantum vorticity, similar to the formation of Abrikosov vortices in type-II superconductors in external magnetic field.

To construct our flux tube model we need another ingredient, which is an abelian gauge field from the decomposition of the gluon potential \cite{Cho, DuanGe, Faddeev, Kondo14} 
\begin{equation}
A_{\mu} = \tilde{A}_{\mu} + B_{\mu}, ~~
\tilde{A}_{\mu} = (A_{\mu} \cdot {\bf n}) {\bf n} + i g^{-1} [ {\bf n}, \partial_{\mu} {\bf n} ], ~
B_{\mu} = i g^{-1} [ \tilde{\nabla}_{\mu} {\bf n}, {\bf n}]
\end{equation}
where $\tilde{\nabla}_{\mu} {\bf n} = \partial_{\mu} {\bf n} - i \, g [A_{\mu}, {\bf n}]$ and a unit vector ${\bf n}$ in the SU(2) color space. Consequently, the field strength can be decomposed into \begin{equation}F_{\mu\nu} = F^{\parallel}_{\mu\nu} + F^{\perp}_{\mu\nu}, ~~
F^{\parallel}_{\mu\nu} = (F_{\mu\nu} \cdot {\bf n} ) {\bf n}, ~~
F^{\perp}_{\mu\nu} = \tilde{\nabla}_{\mu} B_{\nu} - \tilde{\nabla}_{\nu} B_{\mu} \end{equation} 

The original QCD action splits correspondingly. As pointed out in Refs \cite{Kondo14, Cho99}, one of the interesting properties of this decomposition is that $\tilde{A}_{\mu}$ alone is responsible for the Wilson loop and the Polyakov loop at the operator level, i.e.$\mathcal{W} [A] = \mathcal{W}[\tilde{A}], ~\mathcal{L}[A] = \mathcal{L}[\tilde{A}]$ and this is dubbed as the $\tilde{A}$-dominance \cite{Cho99, Kondo14}. This motivates us to use the vector potential $\tilde{A}_{\mu}$ to construct the flux tube for the confinement purpose. We then do a further decomposition
\begin{equation}
\tilde{A}_{\mu} = \Omega^{\textrm{\tiny{vac}}}_{\mu} + Z_\mu {\bf n} 
\end{equation}
where $\Omega^{\textrm{\tiny{vac}}}_{\mu}$ represents a classical QCD vacuum potential and $Z_{\mu}$ is the abelianized gluon potential we need. Its abelian feature can be seen from the field strength 
\begin{equation} \label{CHmunu}
\tilde{F}_{\mu\nu} \equiv   \partial_{\mu} \tilde{A}_{\nu} - \partial_{\nu} \tilde{A}_{\mu} - i \,g [\tilde{A}_{\mu}, \tilde{A}_{\nu}]= Z_{\mu\nu} \,{\bf n},~~~~ Z_{\mu\nu} \equiv \partial_\mu Z_\nu - \partial_\nu Z_\mu
\end{equation}
has an Abelian feature as expected \cite{Cho99}. Now the Yang-Mills action becomes
\begin{equation} \label{YM}
S_{\textrm{YM}} = \int d^4 x \big[ - \frac{1}{4} Z_{\mu\nu}^2  - \frac{1}{4} B_{\mu\nu}^2 -\frac{1}{2} B^{\mu} Q_{\mu\nu} B^{\nu} \big]   
\end{equation}
where the kinetic term of the Abelianized gluon potential $Z_\mu$ has been separated from the other part.
Integrate out the $B_{\mu}$ potential \cite{Kondo14}
\begin{equation}
S_{\textrm{\tiny{gNJL}}} = \int d^4x \bigg( \bar{\psi} ( i \gamma^{\mu} \tilde{\nabla}_{\mu} - \mathcal{M}_Q ) \psi  +\int d^4 y \,G(y) \big[ \bar{\psi}(x+y) \Gamma_{A} \psi(x - y)~\bar{\psi}(x-y) \Gamma_{A} \psi(x + y) \big] \bigg) 
\end{equation}
With local approximation and the usual bosonization procedure for the Nambu-Jona-Lasinio (NJL) Lagrangian, it can be parametrized as
\begin{equation}  \label{eff}
\mathcal{L}_{\textrm{\tiny{eff}}} = \bar{\psi} \big[ i \gamma^{\mu}(\partial_{\mu} - i g Z_{\mu}) - G ( \sigma + i \gamma^5 \pi_a \tau_a ) + \cdots \big] \psi \end{equation} 
where the $\sigma$ term and $\pi$ terms represent the scalar channel and the pseudoscalar channel respectively. The ellipsis includes contributions from the vector and pseudovector channels. This is quite similar to the usual NJL formulation except the gauged piece ($Z_{\mu}$ term), and might be applied to other situations such as quantum cosmology of early universe along the line of Ref.\cite{Channuie}. Noticing that
\begin{equation}
\phi \sim \sigma + i \vec{\pi} \cdot \vec{\tau} = f e^{i \alpha},
\end{equation}
now we can write down a Lagrangian to describe our flux tube model, 
\begin{equation} \label{preNJL}
\mathcal{L}_{\textrm{eff}} = -\frac{\chi(\phi, \phi^*)}{4} Z_{\mu\nu} Z^{\mu\nu} + \partial_{\mu} \phi \partial^{\mu} \phi^* - V(\phi, \phi^*)   - j_{\mu} K^{\mu} 
+ \bar{\psi} \big[ i \gamma^{\mu}(\partial_{\mu}  - i g Z_{\mu}) - g_Y ( \textrm{Re}\phi +  i \gamma^5 \textrm{Im}\phi ) + \cdots \big]\, \psi. 
\end{equation}
The function $\chi(\phi, \phi^*)$ describes the color electric and magnetic polarization properties of the vacuum as a physical medium. The current
\begin{equation}
j_{\mu} \equiv  -\frac{i}{2 |\phi|^2}\left( \phi^{\ast}\partial_{\mu}\phi-\phi\partial_{\mu}\phi^{\ast}\right) =  \partial_{\mu} \alpha
\end{equation}
represents a ``supercurrent" in the quark condensate and the topological current
\begin{equation}
K^{\mu} = \epsilon^{\mu\nu\rho\tau} Z_{\nu} Z_{\rho\tau}
\end{equation}
is the Chern-Simons current. Note that if the $\phi$ field, as the order parameter for the quark condensate, also describes its ``superfluidity", then the current $j_{\mu}$ is actually related to the superfluid velocity. One may ask if the phase of the $\phi$ field can be removed by some rotation of the quark phases. This cannot be done if the $(\phi, Z_\mu, \psi)$ system has nontrivial topological configurations. The phase of the $\phi$ field could be very complicated (see the Fig. 3 in Ref.\cite{XC2} for an example of the phase distribution of a vortex lattice) that it cannot be rotated away due to the topological obstruction. Nevertheless if the phase distribution is topologically trivial it is possible for the $\phi$ field to be real.
In that case we set $g_Y =0$, the current $j_\mu$ vanishes and the couplings $ j_{\mu} J^{\mu}$ and  $ j_{\mu} K^{\mu}$ drop out. The Lagrangian (\ref{preNJL}) reduces to 
\begin{equation} \label{KS}
\mathcal{L}_{\textrm{KS}} = -\frac{1}{4} \chi(\phi) ~Z_{\mu\nu} Z^{\mu\nu} + \frac{1}{2} \partial_{\mu} \phi \partial^{\mu} \phi - V(\phi)  + \bar{\psi} \big[ i \gamma^{\mu}(\partial_{\mu}  - i g Z_{\mu})  \big]\, \psi 
\end{equation}
which is the original flux-tube model constructed by Kogut and Susskind in Ref.\cite{KS75} except that the $Z_\mu$ field here is from the decomposition described at the beginning of this section.

\section{Vortex configuration and emergent topological current term}
\label{sec-3}

In this section we show that if we assume that the QCD string or flux tube induce a vortex configuration of the field $\phi$ in the quark condensate, then a topological current (or topological charge) term will emerge as an effective interaction. First we need check the boundary condition of $\phi$. In order to describe, for instance a single straight vortex-line configuration with winding number $m \in \mathcal{Z}$, one takes the amplitude and the phase of  $\phi$ to be, respectively,
\begin{equation} \label{alpha-theta}
f = f(\rho), ~~~~\alpha = m \theta, ~~ m = \pm 1, \pm 2, \cdots 
\end{equation}
with boundary condition 
\begin{equation} \label{vortexbc}
f (0) = 0, ~~~~\textrm{and} ~~ f(\infty) \longrightarrow \textrm{constant},
\end{equation}
where $\rho$ and $\theta$ are the polar coordinates describing the cross section of the straight vortex line.  The condition on $f(\infty)$ can come readily from the quark condensate value, and the question is whether $f (0) = 0$ can be satisfied inside the flux tube.  Interestingly this has been studied before in the context of the chiral symmetry restoration \cite{Suganuma}. Based on the effective potential 
\begin{equation} 
\mathcal{V}_{\textrm{eff}} ( \sigma, \pi) = - \frac{1}{4 G} ( \sigma^2 + \pi^2) - \frac{i }{2} \textrm{Tr} \ln \bigg( (\partial_{\mu} - i g Z_{\mu}) ^2 
- \frac{g}{2} \sigma_{\mu\nu} Z^{\mu\nu} + \sigma^2 + \pi^2 - i\epsilon \bigg),
\end{equation} 
the result is that when the chromoelectric field exceeds some critical strength,  (see e.g. a numerical study by Suganuma and Tatsumi \cite{Suganuma}:  $E_{crit} \sim 4 $GeV/fm, $E \approx 5.3 $ GeV/fm $>E_{crit}$ ), the chiral symmetry is restored inside the QCD string. The physical picture is simple -- the strong chromoelectric field in the flux tube pulls quark and anti-quark pairs apart and hence, destroys the quark condensate.  
Therefore it is possible that the quark condensate vanishes inside the flux tube, $\langle \bar{q}q \rangle = 0 $, while outside the flux tube the QCD vacuum has non-vanishing $\langle \bar{q}q \rangle \neq 0 $. This shows the quark condensate has the boundary condition of a  vortex configuration. We then assume that the vortex configuration is energetically favored, following the appearance of the Abrikosov vortices in the type-II superconductors in an external magnetic field. Now 
the Dirac equation (taking $ Z_{\mu} = (Z_0, Z_1, 0, 0)$
\begin{eqnarray}  
i \gamma^i (\partial_i - ig Z_i) \psi_L + i( \gamma^2 \cos \theta  + \gamma^3 \sin \theta ) \partial_{\rho} \psi_L  &=& -  f(\rho) e^{- i \theta} \psi_R , \cr
i \gamma^i (\partial_i - ig Z_i) \psi_R + i( \gamma^2 \cos \theta  + \gamma^3 \sin \theta ) \partial_{\rho} \psi_R  &=& - f(\rho) e^{+ i \theta} \psi_L, 
\end{eqnarray} 
from which Callan and Harvey have shown that there exist localized quark zero-modes, more precisely the quark zero-mode has an exponential profile
\begin{equation}\label{4Dpsi}
\psi_L = \chi_{L} \, \exp \big[- \int_0^{\rho} f(\rho') d\rho' \big] \end{equation} 
where the spinor $\chi_{L}$ satisfies a two-dimensional Dirac equation $i \gamma^i (\partial_i - ig Z_i) \chi_{L} = 0$. (It is interesting to compare this localization of chiral zero modes in the flux-tube with the instanton case. 't Hooft has shown that chiral zero modes are located in the instanton \cite{tHooft76} with a profile $\lambda^{3/2}/[(x-a)^2+\lambda^2 ]^{3/2}$ ($\lambda$ and $a$ are the instanton parameters).  This localization of quark zero-modes has a remarkable consequence. These zero-modes are coupled to the gauge potential $Z_i (i=0, 1)$ which leads to a gauge anomaly on the string 
\begin{equation}
\mathcal{D}^k J_k = \frac{1}{2 \pi} \epsilon^{ij} \partial_i Z_j, ~~~ i, j, k = (0, 1).
\end{equation}
As Callan and Harvey have shown in their anomaly-inflow picture \cite{Callan-Harvey}, this gauge anomaly can be cancelled out by an effective action in the bulk space
\begin{equation} \label{cseff}
S_{\textrm{\tiny{C-S}}} = -\frac{ g^2 N_f}{16 \pi^2}\int d^4 x \, \partial_{\mu} \theta \, K^{\mu} \longrightarrow   -\frac{ g^2 N_f}{16 \pi^2}\int d^4 x \, \partial_{\mu} \alpha \, K^{\mu} \sim \int d^4 x \, j_{\mu} \, K^{\mu}
\end{equation}
which is exactly the new topological coupling in the Lagrangian (\ref{preNJL}). Note that $\alpha =  \theta$ since we are considering a static and straight flux tube. To describe a dynamical flux tube that could rotate, vibrate, or even decay into particles, $m \theta$ should be promoted to a spacetime-dependent phase $\alpha(\vec{x}, t)$. This cancellation happens  because the massive quark modes which live off the vortex mediate an effective interaction between the quark condensate and the gluon field, which induces a vacuum current (see Fig. 1) \cite{Callan-Harvey}
\begin{equation}
J^{\textrm{\tiny{ind}}}_{\mu} =  \frac{g^2 N_f}{8 \pi^2} \epsilon_{\mu\nu\rho\tau} Z^{\rho\tau} \partial^{\nu} \theta.
\end{equation}
Converting it to an effective action we obtain (\ref{cseff}). Therefore a topological term which reflects the axial anomaly can be derived from the axionic QCD string model, which is not a pure Yang-Mill configuration like the instanton. The mathematical structure of the anomaly-inflow mechanism is the Stora-Zumino descent equation $\delta \mathcal{K}^{i-1}_{2n-i} = d\, \mathcal{K}^{i}_{2n-i-1} $ \cite{Stora}
\begin{equation} \label{descent}
\delta \int_{M} d\theta \wedge \mathcal{K}^0_{3} = \int_{M} d\theta \wedge d \mathcal{K}^1_{2} 
= - \int_{M} d^2\theta \wedge  \mathcal{K}^1_{2} 
= - \int_{\Sigma} \mathcal{K}^1_{2}
\end{equation}
where $ \mathcal{K}^0_{3}$ is the Chern-Simons 3-form and connected to $\mathcal{K}^1_{2} $ through the descent equation.
Eqt. (\ref{descent}) relates the four dimensional axial anomaly $d \mathcal{K}^0_{3} $ to the lower-dimensional gauge anomaly $ \mathcal{K}^1_{2} $ on the topological defect $\Sigma$.

 \begin{figure}
\centering
 \includegraphics[scale=0.3]{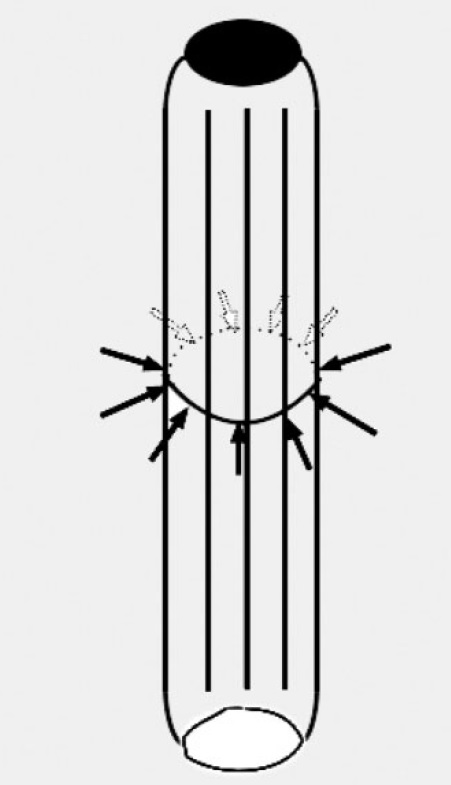}
\caption{``Charging an anomaly battery" -- The anomaly-inflow picture of a chromoelectric flux tube in the QCD vacuum --- The black circle and the white circle represent color sources (e.g. a quark-antiquark pair) and the lines between them are chromoelectric field lines squeezed into a thin tube in the vacuum (shaded area). The arrows on the circle indicate the directions of the induced vacuum current on a particular cross-section of the flux tube, similar to the configuration of the Hall effect \cite{XC}.} 
   \end{figure} 

%
\section{The $U_A(1)$ problem}
\label{sec-4}

With the effective topological current term we can address the $U_A(1)$ problem in different approaches which have been studied before, for example the anomalous Ward identities (ghost pole) \cite{Crewther, Luscher}, the large-$N_c$ limit \cite{Witten, Veneziano} and the 't Hooft vertex in the Nambu-Jona-Lasinio model \cite{tHooft76}. We briefly mention how these methods still work in our formulation: 
\begin{itemize}
\item I. Anomalous Ward identities and ghost pole. Note that the Chern-Simons current $K_{\mu}$ is a long range field
\begin{equation}
\langle
K_{\mu}(x) K_{\nu}(y)\rangle_{q \rightarrow 0} = c_1 \frac{q_{\mu} q_{\nu}}{q^4} + c_2 \frac{g_{\mu\nu}}{q^2}
\end{equation}
With this pole one may avoid the incorrect mass of the $\eta'$ meson from the anomalous Ward identities \cite{Crewther, Lusher}; 
\item II. Large-N approach \cite{Witten, Veneziano}. The usual chiral field should be parametrized as
\begin{equation} 
\tilde{U} = \langle \bar{q} q\rangle \exp{[i \alpha + i \sqrt{2}/F_{\pi} ( \pi_a \lambda_a + \eta_0 /\sqrt{N_f})}]
\end{equation}
 The effective Lagrangian in the large-N limit 
\begin{equation} 
\mathcal{L}_{N} = - \alpha q(x) + \frac{i}{2} q(x) \textrm{Tr}(\log U - \log U^{\dagger}) + \frac{N}{a F^2_{\pi}} q^2(x) + \cdots
\end{equation}
where $q(x)$ is the topological charge density and the ellipsis include other non-topological terms. The usual $\theta$ parameter is replaced by the phase of the quark condensate $\alpha$, which can not be rotated away due to the topological obstacles, then eliminate the field $q(x)$ through its equation of motion and continue the rest of calculations to get the correct mass term for the $\eta'$-meson; 
\item III. The  't Hooft vertex (or the 't Hooft determinant term) in the Nambu-Jona-Lasinio model \cite{tHooft76, Hatsuda}. Noticing that the instanton configuration leads to an effective fermion interaction vertex, so-called  't Hooft vertex, one may ask whether the flux-tube configuration might lead to a similar effective vertex, due to the localization of chiral zero modes (see the comparison of two different localization scenarios in Sec.\ref{sec-4}. We use the general method of Creutz \cite{Creutz2007} which includes fermionic source terms into the partition functional
\begin{equation}
Z (\eta, \bar{\eta}) = \int [\mathcal{D}A]  [\mathcal{D}\bar{\psi}] [\mathcal{D} \psi] e^{ - S_A - (\bar{\psi}, ~(\slashed{D}+m)  \psi) - (\bar{\psi}, ~\eta) - (\bar{\eta},~ \psi)}
 = \int [\mathcal{D}A] e^{ - S_A +~ (\bar{\eta}, ~(\slashed{D}+m)^{-1}  \eta)} \prod_i (\lambda_i + m) 
\end{equation}
where $\lambda_i$ are eigenvalues of the operator $\slashed{D}$ in a flux-tube background, and $m$ is a small explicit mass introduced for the infrared issue of massless quarks. Since the quark sources have overlaps with the chiral zero mode, for instance, $(\bar{u} \cdot \psi_{0L}) \neq 0$, a factor of $m^{-1}$ in the source term cancels a factor of $m$ from the determinant part, so the QCD string configuration can contribute to the correlation functions. Consequently an effective interaction $\sim (\bar{u} \cdot \psi_{0L}) (  \psi^{\dagger}_{0L} \cdot u) \sim \bar{u} (1 \pm \gamma_5 )u $ is obtained. For $N_f \geq 2$ flavors,  each flavor has to contribute for the cancellation of mass factors. This leads to an effective $2 N_f$ vertex which is in the form of a determinant, i.e. the 't Hooft vertex
\begin{equation}
\mathcal{L}_{\textrm{tHooft}} \sim \textrm{det}[ \bar{\psi_i} (1- \gamma_5) \psi_j] + h.c. 
\end{equation}
which is added to the NJL effective Lagrangian (\ref{eff}). Then it has no problem in addressing the U(1) problem and produces the correct mass for the $\eta'$ meson \cite{Hatsuda}. 
\end{itemize}

\section{Chiral magnetic effect and other applications}
\label{sec-5}

The topological defects of the quark condensate might also be induced by other factors, for example, rotations and collisions (see \cite{XC2} for a numerical simulation on rotating condensates). It is well known that a vortex lattice will arise in a rotating superfluid, like liquid Helium II and Bose-Einstein condensates, when the angular velocity is above some critical velocity. For ultra-relativistic collision of heavy nuclei, net color charges develop and flux tubes, with both color electric and magnetic field ($\vec{E} \cdot \vec{B} \neq 0)$, connect the color charges. Therefore longitudinal color electric and magnetic fields can be produced after high energy hadronic collisions. Such gauge configurations carry topological charges and lead to net chiral or antichiral quark zero modes in the flux tubes (see Fig 2. for a couple of examples including quarks hopping on (anti)instantons) \cite{Schafer:1996, XC1}. An interesting phenomenon is that they can separate charge in a background (electromagnetic) magnetic field, and consequently, an electromagnetic current is generated along the magnetic field. This is called the chiral magnetic effect \cite{Kharzeev07, Fukushima08}. It is easy to see that we should have an extra effective coupling \cite{XC}
\begin{equation} \label{mcs}
\Delta\mathcal{L}_{\textrm{\tiny{MCS}}} \sim -\frac{e^2}{8 \pi^2} \, \partial_{\mu} \alpha \, K^{\mu}_{\textrm{\tiny{MCS}}}
\end{equation}
where $ K^{\mu}_{\textrm{\tiny{MCS}}} =  \epsilon^{\mu\nu\rho\tau} A_{\nu} F_{\rho\tau} $ is the Maxwell-Chern-Simons current. This leads to a modified Maxwell equation
\begin{equation}
\partial_{\mu} F^{\mu\nu} = J^{\nu} - \frac{e^2}{2 \pi^2} \, \partial_{\mu} \alpha \, \tilde{F}^{\mu\nu}
\end{equation}
and the induced current $\vec{J} \sim -\dot{\alpha} \vec{B}^{\textrm{\tiny{M}}}$ agrees with \cite{Kharzeev07, Fukushima08}.

 \begin{figure}
 \centering
   \includegraphics[height=7.0cm]{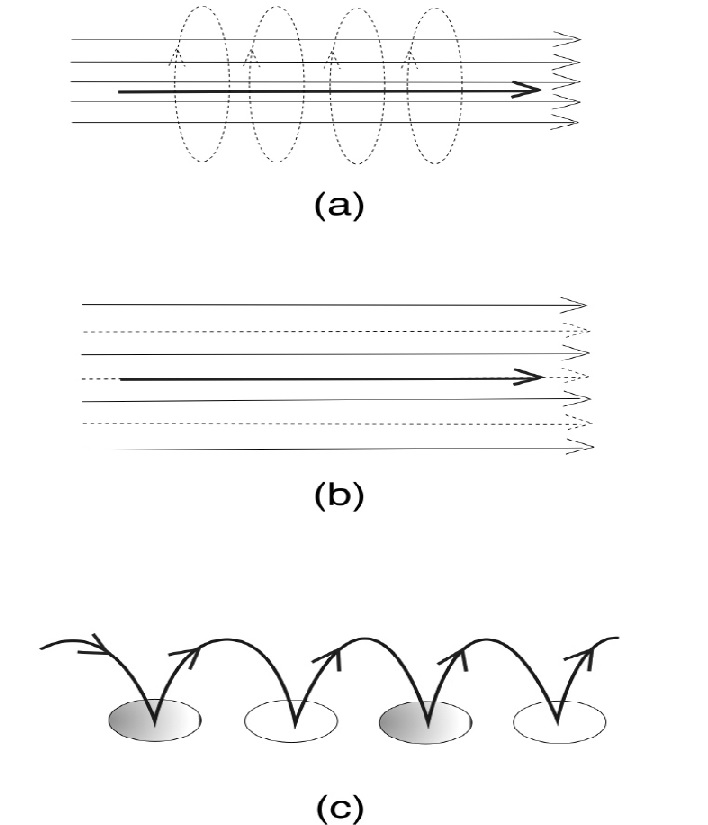}
    \caption{A sketch of different gauge configurations and localizations of chiral zero modes of quarks \cite{XC1} (a) color electric flux-tube; (b) collinear color- electric and magnetic flux-tube; (c) a flowline in the instanton liquid \cite{Schafer:1996} where quarks travel by hopping on the instantons and the anti-instantons. The solid thick lines represent the motion of the quarks. The thin solid lines and dotted lines represent color electric and magnetic fields respectively. The grey circles and white circles are instantons and anti-instantons respectively.}
   \end{figure} 

%

\section{Discussions and conclusions}
\label{sec-6}

In this presentation we give a summary on a new formulation of the QCD string/flux tube model and its applications in the $U_A(1)$ problem and other cases. The main result is that based on this string/flux tube model we can derive a different effective topological term in QCD, i.e. 
\begin{equation}  ~~~\theta \,F \tilde{F}  \longrightarrow  \partial_{\mu}\alpha \, K^{\mu}~~~ \end{equation} 
and other applications follow from that.

We close by a discussion on how the strong CP problem might be avoided in our formulation. Notice that the topological term $\partial_{\mu}\alpha \, K^{\mu}$ is a derivative coupling which is invariant under a constant phase shift $\alpha \rightarrow  \alpha + const$. This shares some common features with the axion approach in solving the strong CP problem. However, in our formulation we do NOT introduce any new particle like the axion, whose role seems to be played by the phase of the quark condensate.  The anomaly inflow mechanism plays an important part in our formulation. It reminds us a question from T. D. Lee on the missing symmetry and quantum numbers \cite{TDLee}: will the physical vacuum contain all those missing symmetries, such that we can restore them if we consider both the matter system and the vacuum?  for example, is it possible that
\begin{equation}
\frac{d}{dt} \left[\begin{array}{c}
P \\ 
C\\ 
CP
\end{array} \right]_{\textrm{matter}} \neq 0  \longrightarrow \frac{d}{dt} \left[\begin{array}{c}
P \\ 
C\\ 
CP
\end{array} \right]_{\textrm{matter+vacuum}} = 0 ?
\end{equation}
In the QCD case, we may expect symmetry conservations
\begin{eqnarray}
\delta|_{\textrm{\tiny{gauge}}} ( \partial_{\mu}\alpha \, K^{\mu}+ \textrm{QCD string}) &=& 0, \cr
\delta|_{\textrm{\tiny{CP}}} ( \partial_{\mu}\alpha \, K^{\mu} + \textrm{QCD string}) &=& 0.  
\end{eqnarray}
and we have shown that the first equation can be realized and leave the second to future investigation. 

\vspace{0.5cm}

{\bf Acknowledgments} I thank  Hank Thacker, Peter Minkowski and Michael Creutz for valuable discussions. 
%
%
%

\end{document}